\documentstyle[11pt,twoside,jltp,epsfig]{article}

\title{Pseudogap and quantum-transition phenomenology in HTS cuprates }
\author{J.L. Tallon$^1$, J.W. Loram$^2$ and C. Panagopoulos$^2$
\footnote{Corresponding author.~E-mail: j.tallon@irl.cri.nz}}
\address{
1. MacDiarmid Institute for Advanced Materials and Nanotechnology,
Industrial Research Ltd. and Victoria University of Wellington,
P.O Box 31310, Lower Hutt, New Zealand\\
2. IRC in Superconductivity and Cavendish Laboratory, Cambridge
University, Cambridge CB3 0HE, England}

\runninghead{Tallon, Loram and Panagopoulos}{ Pseudogap and
quantum transition phenomenology }

\begin{document}

\begin{abstract}
The low-energy excitation spectrum of HTS cuprates is examined in
the light of thermodynamic, transport, quasiparticle and spin
properties. Changes in the thermodynamic spectrum associated with
the normal-state pseudogap disappear abruptly at a critical doping
state, $p_{crit}$ = 0.19 holes per Cu. Moreover, ARPES data at
100K show that heavily damped quasiparticles (QP) near ($\pi$,0)
suddenly recover long lifetimes at $p_{crit}$, reflecting an
abrupt loss of scattering from AF spin fluctuations. This picture
is confirmed by $\mu$SR zero-field relaxation measurements which
indicate the presence of a novel quantum glass transition at
$p_{crit}$. Consistent with this picture resistivity studies on
thin films of Y$_{0.7}$Ca$_{0.3}$Ba$_2$Cu$_3$O$_{7-\delta}$ reveal
linear behavior confined to a V-shaped domain focussed on
$p_{crit}$ at $T$=0. The generic phase behavior of the cuprates
may be governed by quantum critical fluctuations above $p_{crit}$
and the pseudogap appears to be caused by short-range AF
correlations.

\end{abstract}

\maketitle

\section{INTRODUCTION}
The highly anomalous properties of the high-$T_c$ superconducting
(HTS) cuprates are attributable in part to the presence of a
normal-state (NS) pseudogap extending across the underdoped region
of the phase diagram. The onset of the pseudogap opens up a gap in
the density of states (DOS) which profoundly affects all physical
properties in both the normal and superconducting (SC)
states\cite{cooper}. Considerable debate has ensued as to the
nature of the pseudogap and, in particular, whether it is some
form of precursor SC pairing\cite{emery} or a NS correlation which
competes (or coexists) with SC\cite{loram1}. We believe the
evidence is firmly in support of the latter scenario and present
here further supporting evidence. In particular we show that the
SC ground state is strongly perturbed by the pseudogap and that
the magnetic spectrum undergoes abrupt and major changes at the
critical doping state $p_{crit}$ at which the pseudogap first
appears. (Here $p$ is the doped hole concentration per planar Cu).
Finally, we show from thin-film transport measurements that the
famous linear resistivity is confined to a V-shaped region in the
$T$-$p$ phase diagram that is centered on $p_{crit}$. That is,
only at $p_{crit}$ does the NS resistivity remain linear to $T=0$.
This behavior is consistent with a quantum critical point (QCP)
scenario\cite{tallon1}.

\section{THERMODYNAMIC PROPERTIES}
Our starting point is to recognize that the thermodynamic and
transport behavior discussed here seems to be universal amongst
the HTS cuprates. Thus we will alternate between the three
compounds Y$_{0.8}$Ca$_{0.2}$Ba$_2$Cu$_3$O$_{7-\delta}$ (Y-123),
Bi$_2$Sr$_2$CaCu$_2$O$_{8+\delta}$ (Bi-2212) and
La$_{2-x}$Sr$_x$CuO$_4$ (La-214) under the assumption that our
deductions are universal. Of course there are some features which
seem unique to La-214, such as a strong stripe character and a
high degree of disorder, but the properties we discuss here are
also realized in the other two systems.

Each of these systems exhibits abrupt changes in their
thermodynamic properties at $p_{crit}$. This is illustrated, in
the case of Bi-2212, in Fig.~1 using previously reported
data\cite{loram2}. Panel (a) shows the doping dependence of the
jump in the specific heat coefficient, $\Delta\gamma$, and of the
pseudogap energy, $E_g$, obtained by fitting the temperature
dependence of the entropy using a triangular NS DOS
$N(E)=N_0(E/E_g)$ for $E<E_g$ and $N(E)=N_0$ for $E>E_g$. Such a
DOS is non-states conserving and has been shown to fit the
thermodynamic data very well over a broad range of $T$ and
$p$\cite{loram2}.

\begin{figure}
\centerline{\includegraphics*[height=85mm,
width=75mm]{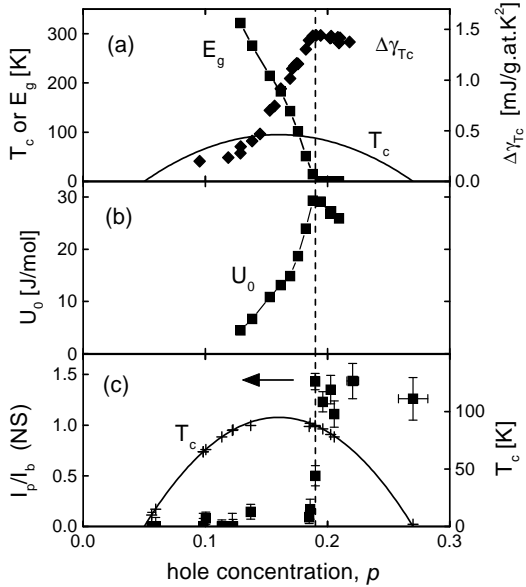}} \caption{\small  The doping dependence
of various parameters for Bi-2212 ($a$) $\Delta\gamma$ and $E_g$,
($b$) $U_0$ and ($c$) the normalized quasiparticle peak at
($\pi$,0) and 100 K.}
\end{figure}

Panel (b) shows the $p$-dependence of the condensation energy
$U_0$. Notably the opening of the pseudogap, as shown by the value
of $E_g$, is abrupt and it has an immediate effect in sharply
depressing $U_0$ and $\Delta\gamma$ (a measure of the pair
density). Accompanying these abrupt changes the superfluid density
at $T=0$ is also sharply reduced, as shown by ac
susceptibility\cite{panag1} field-dependent specific heat
measurements\cite{loram2} and $\mu$SR\cite{bernhard}. This
reflects a sudden change in the ground state with the opening of
the pseudogap and is not compatible with the precursor pairing
models but rather suggests that the pseudogap represents some
correlated state that coexists with SC.

\section{QUASIPARTICLE WEIGHT AND LIFETIME}
The Bi-2212 system has been widely investigated using other
techniques, and many key observations on the pseudogap have been
assembled for this system. We mention briefly recent scanning
tunnelling microscopy (STM) measurements which appear to suggest a
very high degree of local inhomogeneity. Pan {\em et
al.}\cite{pan} suggest that such disorder may be a characteristic
of all HTS systems. We believe this not to be the case.
Inhomogeneity involving separate spatial regions of SC and NS
metallic behavior would result in a finite DOS at the Fermi level
that is not observed, while spatially separate insulating and
metallic regions would result in a reduction of the NS DOS at high
temperature, $T > E_g/k_B$. This is not observed either, indeed
the specific heat coefficient for Bi-2212 is constant at high $T$,
independent of doping\cite{loram2}. A stronger case, still, can be
made with reference to NMR linewidths and transition broadening.
Moreover, the abruptness of the various onsets shown in Fig. 1
raise major questions about the very large static spatial
inhomogeneity inferred from STM ($\pm$30\% spread in energy gap,
local DOS and doping level). We emphasize that the heat capacity
is a bulk measurement while STM probes just the outermost CuO$_2$
layer which is likely to be subject to inhomogeneity arising from
reconstruction, oxygen disorder and Bi/Sr intersubstitution.

Another technique that has contributed in a major way to our
understanding of the cuprates is angle-resolved photoelectron
spectroscopy (ARPES). Systematic studies of underdoped samples
show that NS quasiparticle (QP) lifetimes are very short near
($\pi$,0) due to pronounced scattering from antiferromagnetic (AF)
spin fluctuations because it is at the zone boundary that the AF
wave vector spans the Fermi surface. This is manifested by the
almost complete absence of the NS QP peak\cite{arpes1} whose
weight is redistributed due to these scattering events. On the
other hand, near the zone diagonal QP lifetimes are long and the
QP peak is fully recovered\cite{arpes1}. We have analyzed the
leading-edge energy dispersion curves for a large number of
reported measurements on Bi-2212
\cite{arpes1,arpes2,arpes3,arpes4,arpes5,arpes6,arpes7,arpes8} at
($\pi$,0) and at 100K (i.e. in the NS) and we have fitted these to
a background curve plus a Gaussian QP peak. The magnitude of the
latter has been ratioed with that of the former (as a measure of
the loss or redistribution of QP weight) and plotted as a function
of doping in panel ($c$) of Fig. 1. Remarkably the QP weight is
very small across the entire underdoped region and is abruptly and
fully recovered at $p_{crit}$ in the lightly overdoped region.
(The most highly doped sample is for Bi-2201). While the data used
is obtained from a variety of sources with differing experimental
conditions, the changes are so marked that they must reflect a
real and dramatic alteration in the spin spectrum, presumably the
almost complete demise of scattering from AF fluctuations.
Elsewhere we have presented evidence from other studies for the
collapse of short-range AF correlations at $p_{crit}$. This
includes inelastic neutron scattering in Y-123, induced local
moments in Zn-substituted Y-123 samples and two-magnon Raman
scattering in Bi-2212\cite{tallon2}. This accumulated evidence is
rather strong and implies that the pseudogap is associated with
short-range AF correlations which suddenly disappear at
$p_{crit}$, as does the pseudogap.

\section{QUANTUM GLASS TRANSITION}
One of the questions which arises from the foregoing is whether
$p_{crit}$ is a QCP. Thermodynamic measurements show that there is
no phase transition associated with the characteristic temperature
$T^* = E_g/k_B$ which should rather be considered as a cross-over
temperature. Various authors have proposed some form of hidden
order parameter which is established below $T^*$\cite{varma}. None
has thus far been substantiated. However we note that a quantum
phase transition at $p_{crit}$ could be associated with a
spin-glass transition with long-range Edwards-Anderson
order\cite{Edwards}.

We have recently completed a series of zero-field $\mu$SR studies
on the spin fluctuation spectrum in pure and Zn-substituted
La-214\cite{panag2}. These investigations allow the determination
(for any given doping state) of the temperature, $T_f$, at which
spin fluctuations first enter the $\mu$SR time window ($10^{-10}$
sec) and the temperature, $T_g$, at which they leave the $\mu$SR
time window ($10^{-6}$ sec) i.e. at which they become essentially
static as far as the $\mu$SR technique is concerned. These two
crossover temperatures are plotted in Fig.~2 for
La$_{2-x}$Sr$_x$Cu$_{1-y}$Zn$_y$O$_4$ with $y$ = 0, 1 and 2 \%.
For each Zn concentration the pairs of crossover temperatures,
$T_f$ and $T_g$, converge to zero at $p_{crit}$. Thus at
$p_{crit}$ the rate of slowing down of spin fluctuations diverges,
a clear signature of a quantum glass transition. We note, without
further discussion, the enhancement in $T_f$ and $T_g$ seen at
1/8$^{\textrm{th}}$ doping which coincides with the local
reduction in $T_c$. Evidently, AF spin fluctuations slow markedly
in this region, possibly due to incipient stripe formation.

\begin{figure}
\centerline{\includegraphics*[height=80mm,width=90mm]{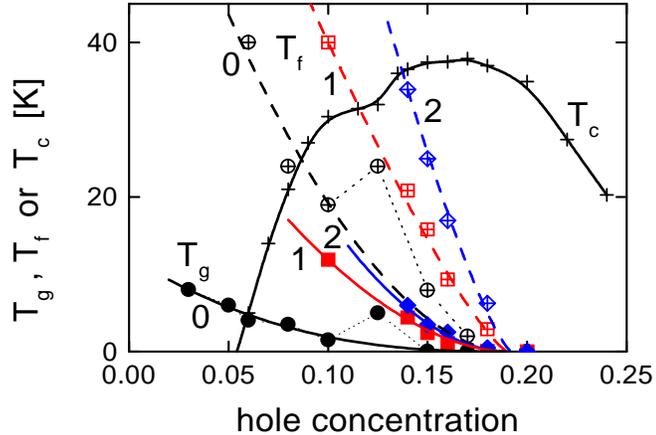}}
\caption{\small  The doping dependence, for La-214, of the upper
(open symbols, dashed curves) and lower (solid symbols, solid
curves) crossover temperatures, $T_f$ and $T_g$, at which spin
fluctuations first enter (10$^{-10}$ sec) and leave (10$^{-6}$
sec) the $\mu$SR time window. Data is shown for 0 (black), 1 (red)
and 2\% (blue) Zn substitution for Cu.}
\end{figure}

AF spin fluctuations are also slowed by Zn substitution and this,
again, moves $T_f$ and $T_g$ to higher values for a given
$p$-value. In each case there is no evidence of slow fluctuations
(within the $\mu$SR time window) for $p > p_{crit}$. One
interpretation is that the mobile carriers for $p
> p_{crit}$ flip spins so rapidly that the AF spin fluctuation lifetime is
suppressed effectively to zero. There emerges, then, the picture
that for $p < p_{crit}$ the AF spin lifetime is long but the QP
lifetime at ($\pi$,0) is strongly attenuated due to scattering
from AF spin fluctuations while for $p > p_{crit}$ the QP lifetime
at ($\pi$,0) recovers while the spin lifetime is strongly
attenuated. The crossover is remarkably abrupt.

\section{LINEAR RESISTIVITY AND THE QCP}
If there is a quantum glass transition (QGT) at $p_{crit}$ then
one expects, within the phenomenology of quantum critical
behavior, a resistivity which is linear in $T$ occurring at
$p_{crit}$\cite{tallon1}. To either side of the QGT the region of
linearity would be expected to be pushed to higher temperatures
thus describing a V-shaped domain of linear-$T$ resistivity in the
quantum critical regime which is focussed on $p_{crit}$. Within
this picture the crossover line $T^*(p)$ would form the boundary
of the "V" for $p < p_{crit}$ while another line $T'(p)$, for $p >
p_{crit}$, would form the other boundary of the "V". This is
precisely what is observed in the three model cuprates we are
discussing. Below $T^*(p)$ the resistivity $\rho(T)$ is found to
be sub-linear while below $T'(p)$ the resistivity $\rho(T)$ is
super-linear. We have previously analyzed the resistivity of
epitaxial thin films of
Y$_{0.7}$Ca$_{0.3}$Ba$_2$Cu$_3$O$_{7-\delta}$ to show this very
behavior\cite{tallon1}. In Fig.~3 we show this data in a more
illustrative way. We plot a map of the $p$- and $T$-dependence of
the resistivity with the high-temperature linear-$T$ dependence
divided out. The figure shows that the region where
$\rho(T)/(\rho_0 +\alpha T)$ is close to unity indeed forms a
V-shape about the critical doping state $p_{crit}$ = 0.19 holes
per Cu. As can be seen, the tendency of this region to focus to
$p_{crit}$ at $T=0$ is cut-off by the appearance of
superconductivity. Ideally one would like to suppress SC so as to
follow this behavior to low temperature. The use of Zn
substitution provides a possible route but the data is complicated
by the effects of residual scattering and localization.
Nonetheless Fig. 3 illustrates a phenomenology strongly
reminiscent of QCP behavior. Similar studies could be carried out
in high pulsed magnetic fields (perhaps combined with Zn
substitution) to expose this behavior to low temperatures.

\begin{figure}
\centerline{\includegraphics*[width=80mm]{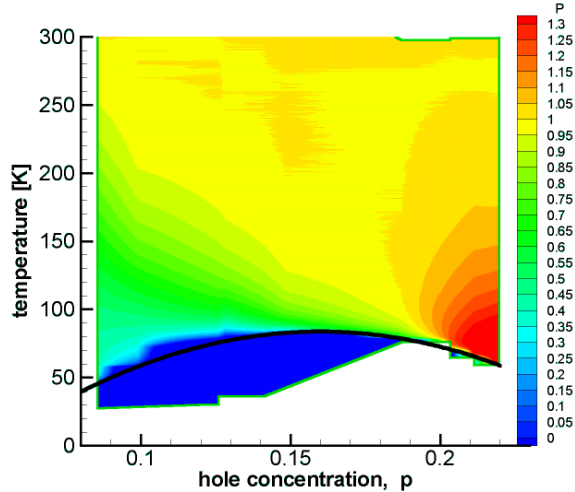}}
\caption{\small  Colour-scale plot, on the $T-p$ phase diagram, of
$\rho(T)/(\rho_0 +\alpha T)$ for the resistivity of thin films of
Y$_{0.7}$Ca$_{0.3}$Ba$_2$Cu$_3$O$_{7-\delta}$. $T_c(p)$ is shown
by the solid curve.}
\end{figure}

\section{CONCLUSION}
In the light of these compelling evidences for coexisting NS
correlations and possible QCP behavior we believe it is time to
abandon the previously widespread notion of the pseudogap deriving
from precursor SC pairing. Clearly such pairing effects occur
close to $T_c$, and indeed have been observed, but the dominant
pseudogap behavior that is observed over a large domain of the
phase diagram (and to high temperature) seems clearly to be
associated with short-range AF spin correlations that result in
the opening of a gap in the NS DOS. This takes away spectral
weight otherwise available for SC pairing around the Fermi surface
with consequent dramatic weakening of the ground-state SC as well
as strongly modifying the NS properties. We hope that the present
data (and other related data) will help establish much-needed
consensus in these matters.

\section*{ACKNOWLEDGMENTS}
Thanks are due to the Marsden Fund of New Zealand (JLT) and the
Royal Society (CP) for financial support for this programme.



\begin{thebibliography}{99}


\bibitem{cooper}  J. R. Cooper and J. W. Loram, J. Phys. I (France) {\bf 6},
2237 (1996).

\bibitem{emery}  V. Emery and S. A. Kivelson, Nature (London) {\bf 374}, 434 (1995).

\bibitem{loram1}  J. W. Loram {\em et al.}, J. Phys. Chem. Solids {\bf 59},
2091 (1998).

\bibitem{tallon1}  J. L. Tallon {\em et al.}, phys. stat. solidi ($b$) {\bf 215},
531 (1999).

\bibitem{loram2}  J. W. Loram {\em et al.}, J. Phys. Chem. Solids {\bf 62},
59 (2001).

\bibitem{panag1} C. Panagopoulos {\em et al.}, Phys. Rev. B {\bf 60},
14617 (1999).

\bibitem{bernhard} C. Bernhard {\em et al.}, Phys. Rev. Lett. {\bf 86},
1614 (2001).

\bibitem{pan}  S. H. Pan {\em et al.}, Nature {\bf 413},
282 (2001).

\bibitem{arpes1}  C. Kim {\em et al.}, Phys. Rev. Lett. {\bf 80},
4245 (1998).

\bibitem{arpes2}  D. S. Marshall {\em et al.}, Phys. Rev. Lett. {\bf 76},
4841 (1996).

\bibitem{arpes3}  R. B. Laughlin, Phys. Rev. Lett. {\bf 79},
1726 (1997).

\bibitem{arpes4}  M. R. Norman {\em et al.}, Phys. Rev. Lett. {\bf 79},
3506 (1997).

\bibitem{arpes5}  Z. -X. Shen and J. R. Schrieffer, Phys. Rev. Lett. {\bf 78},
1771 (1997).

\bibitem{arpes6}  P. J. White {\em et al.}, Phys. Rev. B {\bf 54},
R15669 (1996).

\bibitem{arpes7}  T. Sato {\em et al.}, Phys. Rev. B {\bf 64},
054501-1 (2001).

\bibitem{arpes8}  J. Mesot {\em et al.}, Phys. Rev. B {\bf 63},
224516 (2001).

\bibitem{tallon2}  J. L. Tallon and J. W. Loram, Physica C {\bf 349},
53 (2001).

\bibitem{varma}  C. M. Varma, Phys. Rev. Lett. {\bf 83}, 3538 (1999); S. Chakravarty {\em et al.}, Phys. Rev. B {\bf 63},
10000 (2001).

\bibitem{Edwards}  S.F. Edwards and P.W. Anderson, J. Phys. F: Metal Phys. {\bf 5},
965 (1975).

\bibitem{panag2}  C. Panagopoulos {\em et al.}, Phys. Rev. B {\bf 66},
064501 (2002).


\end{thebibliography}
\end{document}